\documentclass{aastex}
\usepackage{natbib, aas_macros}
\usepackage{graphicx} 
\usepackage{amsmath}
\usepackage{array}
\usepackage{multirow}
\usepackage[mathcal]{euscript}

\begin{document}

\title{Viscoelastic Models of Tidally Heated Exomoons}
\author{Vera Dobos\altaffilmark{1}}
\affil{Konkoly Thege Miklos Astronomical Institute, Research Centre of Astronomy and Earth Sciences, Hungarian Academy of Sciences}
\affil{H--1121 Konkoly Thege Mikl\'os \'ut 15-17, Budapest, Hungary}
\email{dobos@konkoly.hu}
\altaffiltext{1}{Department of Astrophysical Sciences, Princeton University, 08544, 4 Ivy Lane, Peyton Hall, Princeton, NJ, USA}
\and
\author{Edwin L. Turner\altaffilmark{2}}
\affil{Department of Astrophysical Sciences, Princeton University}
\affil{08544, 4 Ivy Lane, Peyton Hall, Princeton, NJ, USA}
\altaffiltext{2}{The Kavli Institute for the Physics and Mathematics of the Universe (WPI), University of Tokyo, 227-8583, 5-1-5 Kashiwanoha, Kashiwa, Japan}

\begin{abstract}

Tidal heating of exomoons may play a key role in their habitability, since the elevated temperature can melt the ice on the body even without significant solar radiation. The possibility of life is intensely studied on Solar System moons such as Europa or Enceladus, where the surface ice layer covers tidally heated water ocean. Tidal forces may be even stronger in extrasolar systems, depending on the properties of the moon and its orbit. For studying the tidally heated surface temperature of exomoons, we used a viscoelastic model for the first time. This model is more realistic than the widely used, so-called {\it fixed~Q} models, because it takes into account the temperature dependency of the tidal heat flux, and the melting of the inner material. With the use of this model we introduced the circumplanetary Tidal Temperate Zone (TTZ), that strongly depends on the orbital period of the moon, and less on its radius. We compared the results with the fixed $Q$ model and investigated the statistical volume of the TTZ using both models. We have found that the viscoelastic model predicts 2.8 times more exomoons in the TTZ with orbital periods between 0.1 and 3.5 days than the fixed Q model for plausible distributions of physical and orbital parameters. The viscoelastic model gives more promising results in terms of habitability, because the inner melting of the body moderates the surface temperature, acting like a thermostat.

\end{abstract}

\maketitle

\section{Introduction}

No exomoons have been discovered yet, but these measurements are expected in the next decade. \cite{bennett14} present a candidate, which has been detected via the MOA-2011-BLG-262 microlensing event. The best-fit solution for the data implies the presence of an exoplanet hosting a sub-Earth mass moon. This measurement however needs confirmation, since an alternate solution is also presented. Nevertheless, this measurement indicates that the era of exomoon detections is about to begin.

The most favorable method for exomoon discoveries is photometry. An exoplanetary transit may enlighten the presence of a moon in the light curve. Details of this method are thoroughly discussed in the literature \citep{simon07, kipping09a, kipping09b, simon10, kipping12, simon12}.

In addition, habitability of exomoons is under examination as well \citep[see e.g.][]{kaltenegger10, heller13, helleretal14}. \cite{hinkel13} investigated the influence of eccentric planetary orbits on moons, and concluded that a moon with sufficient atmospheric heat redistribution may sustain suitable temperature for life on its surface even if it orbits a planet that moves temporarily outside of the habitable zone (HZ) at each orbital period.

Solar System analogs may serve as useful examples for different exomoon types. The satellites in the Solar System are diverse and life on them is a puzzling question. The icy surface of Europa and Enceladus probably covers water ocean, which may provide a suitable environment for life \citep{carr98, kargel00, collins07, iess14}. Tidal and radiogenic heat keeps the interior of the body warm, hence maintain the water in liquid state. In fact, these internal heat sources drive to the eruption of plumes on Enceladus, and similar phenomenon was discovered on Europa as well \citep{porco06, roth14}.

The idea of a circumplanetary, tidally-heated habitable zone has emerged and was investigated by several authors \citep[e.g.][]{reynolds87, scharf06, heller13}. For the first time, we apply a viscoelastic model for studying tidal heat in exomoons. This work aims to give a detailed study of the circumplanetary Tidal Temperate Zone, and discusses the differences with other models.

\section{Viscoelastic model}

\subsection{Advantages}

Tidal heat rate of a moon is usually calculated by the following expression \citep[e.g.][]{reynolds87, meyer07}:

\begin{equation}
   \label{fixQ}
       \dot E_\mathrm{tidal} = \frac {21} {2} \frac {k_2} {Q} \frac {G M_\mathrm{p}^2 R_\mathrm{m}^5 n e^2} {a^6} \, ,
\end{equation}

\noindent where $G$ is the gravitational constant, $M_p$ is the mass of the planet, $R_\mathrm{m}$, $n$, $e$ and $a$ are the radius, mean motion, eccentricity and semi-major axis of the moon, respectively. $Q$ is the tidal dissipation factor and $k_2$ is the second order Love number:

\begin{equation}
   \label{k2}
       k_2 = \frac {3/2} {1 + \frac { 19 \mu } { 2 \rho g R_\mathrm{m} } } \, ,
\end{equation}

\noindent where $\mu$ is the rigidity, $\rho$ is the density and $g$ is the surface gravity of the satellite. This calculation method is called the fixed $Q$ model, because $Q$, $\mu$ and $k_2$ are considered to be constants.

\begin{figure}
\centering   
\includegraphics[width=25pc]{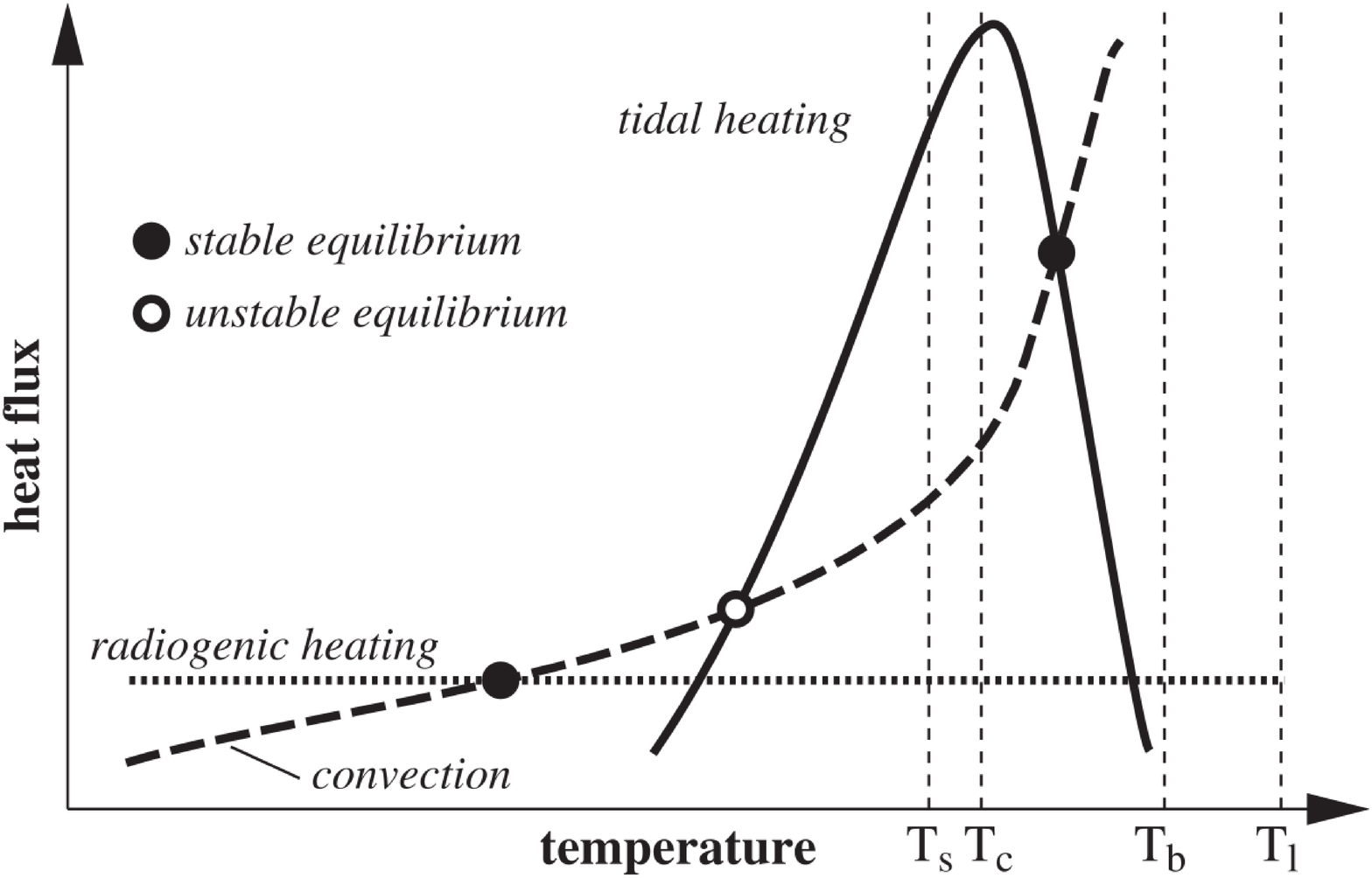}     
\caption{\label{Moore}Schematic figure of the temperature dependency of tidal heat flux and convective heat loss \citep{moore03}.}
\end{figure}

The fixed $Q$ model is broadly used in tidal calculations, but highly underestimates the tidal heat of the body \citep{ross88, meyer07}. Moreover, both $Q$ and $\mu$ are very difficult to determine, and vary on a large scale for different bodies: from a few to hundreds for rocky planets, and tens or hundreds of thousands for giants \citep[see e.g.][]{goldreich66}. In addition, these parameters are not constants, since they strongly depend on the temperature \citep{fischer90, moore03, henning09, shoji14}. As a consequence, tidal heat flux has a temperature dependency, as well: it reaches a maximum at a critical temperature ($T_\mathrm{c}$) as can be seen in Fig. \ref{Moore}. Between the solidus and the liquidus temperature ($T_\mathrm{s}$ and $T_\mathrm{l}$, respectively) the material partially melts. Above the breakdown temperature ($T_\mathrm{b}$) the mixture behaves as a suspension of particles. The dashed curve represents the convective heat loss of the body. Circles indicate equilibria, for example, the solid circle between $T_\mathrm{c}$ and $T_\mathrm{b}$ is a stable equilibrium point. If the temperature increases, convective cooling will be stronger than the heat flux, resulting in a cooler temperature. In case of decreasing temperature, the tidal heat flux will be the stronger, hence the temperature increases, returning the system to the stable point. The stable equilibrium between the tidal heat and convection is not necesserely located between $T_\mathrm{c}$ and $T_\mathrm{b}$, in fact, there are cases, when the two curves do not have intersection at all \citep[see][Fig. 6]{henning09}. In these cases tidal heat is not strong enough to induce convection inside the body.

In contrast to the fixed $Q$ model, viscoelastic models take into account the temperature dependency of the body, hence are more realistic.

\subsection{Description}

In viscoelastic models $k_2/Q$ is replaced by the imaginary part of the complex Love number Im($k_2$), which describes structure and rheology in the satellite \citep{segatz88}:

\begin{equation}
   \label{viscel}
       \dot E_\mathrm{tidal} = - \frac {21} {2} Im(k_2) \frac {R_\mathrm{m}^5 n^5 e^2} {G} \, .
\end{equation}

\noindent Note that in this expression the mass of the planet and the semi-major axis of the moon are eliminated by the mean motion ($n = \sqrt{G M_p / a^3}$).

\citet{henning09} gives the value of Im($k_2$) for four different models (see Table 1. in their paper). In this work we use the Maxwell model:

\begin{equation}
   \label{Imk2}
       - Im(k_2) = \frac {57 \eta \omega} { 4 \rho g R_\mathrm{m} \left[ 1 + \left( 1 + \frac { 19 \mu } { 2 \rho g R_\mathrm{m} } \right)^2 \frac { \eta^2 \omega^2 } { \mu^2 } \right] } \, ,
\end{equation}

\noindent where $\eta$ is the viscosity, $\omega$ is the orbital frequency and $\mu$ is the shear modulus of the satellite.

The viscosity and the shear modulus of the body strongly depend on the temperature. Below the $T_\mathrm{s}$ the shear modulus is constant: $\mu = 50 \, \mathrm{GPa}$ and the viscosity follows an exponential function:

\begin{equation}
   \label{viTs}
       \eta = \eta_0 \, \mathrm{exp} \left( \frac {E} {\mathcal{R} T} \right) \, ,
\end{equation}

\noindent where $\eta_0 = 1.6 \cdot 10^5 \mathrm{Pa \, s}$ (reference viscosity), $E$ is the activation energy, $\mathcal{R}$ is the universal gas constant and $T$ is the temperature of the material \citep{fischer90}.

Between $T_\mathrm{s}$ and $T_\mathrm{b}$ the body starts to melt. The shear modulus changes by

\begin{equation}
   \label{shTb}
       \mu = 10^{ \left( \frac {\mu_1} {T} + \mu_2 \right) } \mathrm{Pa} \, ,
\end{equation}

\noindent where $\mu_1 = 8.2 \cdot 10^4 \mathrm{K}$ and $\mu_2 = -40.6$ \citep{fischer90}. The viscosity can be expressed by

\begin{equation}
   \label{viTb}
       \eta = \eta_0 \, \mathrm{exp} \left( \frac {E} {\mathcal{R} T} \right) \mathrm{exp} \left( -B \phi \right) \, ,
\end{equation}

\noindent where $\phi$ is the melt fraction which increases linearly with the temperature between $T_\mathrm{s}$ and $T_\mathrm{l}$ ($0 \leq \phi \leq 1$) and $B$ is the melt fraction coefficient ($10 \leq B \leq 40$) \citep{moore03}.

At $T_\mathrm{b}$ the grains disaggregate, leading to a sudden drop in both the shear modulus and the viscosity. Above this temperature the shear modulus is set to a constant value: $\mu = 10^{-7} \mathrm{Pa}$. The viscosity follows the Roscoe-Einstein relationship so long as it reaches the liquidus temperature (where $\phi=1$) \citep{moore03}:

\begin{equation}
   \label{viTl}
       \eta = 10^{-7} \mathrm{exp} \left( \frac {40000 \mathrm{K}} {T} \right) \left( 1.35 \phi - 0.35 \right)^{-5/2} \mathrm{Pa \, s} \, .
\end{equation}

Above $T_\mathrm{l}$ the shear modulus stays at $10^{-7} \mathrm{Pa}$, and the viscosity is described by \citep{moore03}

\begin{equation}
   \label{viTll}
       \eta = 10^{-7} \mathrm{exp} \left( \frac {40000 \mathrm{K}} {T} \right) \mathrm{Pa \, s} \, .
\end{equation}

In our calculations rocky bodies are considered as satellites, and for this reason we follow the melting temperatures of \citet{henning09}, namely: $T_s = 1600$~K, $T_l = 2000$~K. We assume that disaggregation occurs at 50\% melt fraction, hence the breakdown temperature will be $T_b = 1800$~K.

\subsection{Internal structure and convection}

The structure of the moon in the model is the following: the body consists of an inner, homogenous part, which is convective, and an outer, conductive layer. If the tidal forces are weak, the induced temperature will be low, resulting in a smaller convective region and a deeper conductive layer. But in case of strong tidal forces, the temperature will be higher, hence the convective zone will be larger with a thinner conductive layer.

For calculating the convective heat loss, we use the iterative method described by \citet{henning09}. The convective heat flux can be obtained from

\begin{equation}
   \label{qBL}
       q_\mathrm{BL} = k_\mathrm{therm} \frac {T_\mathrm{mantle} - T_\mathrm{surf}} {\delta(T)} \, ,
\end{equation}

\noindent where $k_\mathrm{therm}$ is the thermal conductivity ($\sim 2 \mathrm{W/mK}$), $T_\mathrm{mantle}$ and $T_\mathrm{surf}$ are the temperature in the mantle and on the surface, respectively, and $\delta(T)$ is the thickness of the conductive layer. We use $\delta(T)=30 \, \mathrm{km}$ as a first approximation, and then for the iteration

\begin{equation}
   \label{delta}
       \delta(T) = \frac {d} {2 a_2} \left( \frac {Ra} {Ra_\mathrm{c}} \right)^{-1/4}
\end{equation}

\noindent is used, where $d$ is the mantle thickness ($\sim 3000$~km), $a_2$ is the flow geometry constant ($\sim 1$), $Ra_\mathrm{c}$ is the critical Rayleigh number ($\sim 1100$) and $Ra$ is the Rayleigh number which can be expressed by

\begin{equation}
   \label{Ra}
       Ra = \frac { \alpha \, g \, \rho \, d^4 \, q_\mathrm{BL} } { \eta(T) \, \kappa \, k_\mathrm{therm} } \, .
\end{equation}

\noindent Here $\alpha$ is the thermal expansivity ($\sim 10^{-4}$) and $\kappa$ is the thermal diffusivity: $\kappa = k_\mathrm{therm} / ( \rho \, C_\mathrm{p} )$ with $C_\mathrm{p} = 1260 \, \mathrm{J/(kg \, K)}$. For detailed description see the clear explanation of \citet{henning09}.

Because of the viscosity of the material the thickness of the boundary layer and convection in the underlying zone changes strongly with temperature. The weaker temperature dependencies of density and thermal expansivity are neglected in the calculations. The iteration of the convective heat flux lasts until the difference of the last two values is higher than $10^{-10} \mathrm{W/m^2}$.

Calculations of tidal heat flux and convection are made for a fixed radius, density, eccentricity and orbital period of the moon. We assume that with time, the moon reaches the equilibrium state. \citet{henning09} showed that planets with significant tidal heating reach equilibrium with convection in a few million years. However, change in the eccentricity can shift, or destroy stable equilibria. After finding the stable equilibrium temperature, the tidal heat flux is calculated, from which the surface temperature can be obtained using the Stefan-Boltzmann law:

\begin{equation}
   \label{Tsurf}
      T_\mathrm{surf} = \left( \frac{ \dot E_\mathrm{tidal} } { 4 \pi R_\mathrm{m}^2 \sigma } \right)^{1/4}\, ,
\end{equation}

\noindent where $\sigma$ is the Stefan-Boltzmann constant. This is the first time of using a viscoelastic model for obtaining the tidal heat induced surface temperature on exomoons.

\subsection{Results}

The satellite's surface temperature is calculated for different orbital periods and radii, at a fixed density and eccentricity. Stellar radiation and other heat sources are not considered, and have been neglected. The orbital period and the radius of the moon varies between 2 and 20 days, and between 250~km and 6550~km, respectively. It is common to consider Earth-mass moons in extrasolar systems when speaking of habitability, however, their existence is not proven. In the Solar System the largest moon, Ganymede has only 0.025 Earth mass. But the mass of satellite systems is proportional to the mass of their host planet. \citet{canup06} showed that this might be the case for extrasolar satellite systems as well, giving an upper limit for the mass ratio at around $10^{-4}$. This means, that 10 jupiter-mass planets may have Earth-mass satellites. Besides accretion, large moons can also form from collisions, as in the case of the Earth's Moon. Other possibility is the capturing of terrestrial-sized bodies through a close planetary encounter, as described by \citet{williams13}. For these reasons, we also take Earth-like moons into account.

\begin{figure}
	\centering   
	\includegraphics[width=25pc]{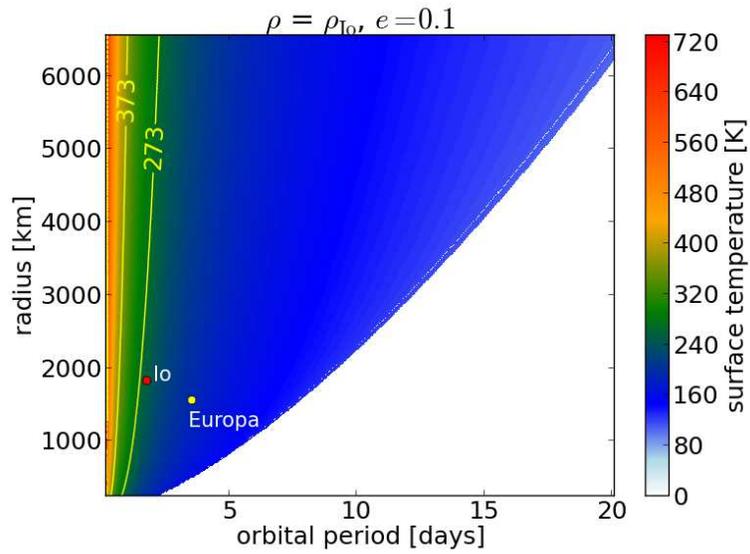}     
	\caption{\label{e01}Tidal heat induced surface temperature for moons with similar density to Io at 0.1 orbital eccentricity.}
\end{figure}

The results can be seen in Fig. \ref{e01}, where the density of the moon is that of Io, and its eccentricity is set to 0.1. Different colours indicate different surface temperatures. In the white region there is no stable equilibrium between tidal heat and convective cooling. In other words, tidal heat is not strong enough to induce convection. For comparison, a few Solar System moons are plotted that have similar densities to Io's. Yellow contour curves denote 0 and $100\,^\circ$C. The green area between these curves indicates that water may be liquid on the surface of the moon (atmospheric considerations were not applied). We define this territory as the Tidal Temperate Zone (TTZ).

Interestingly, the location of the TTZ strongly depends on the orbital period, and less on the radius of the moon. Low radii are less relevant, since smaller bodies are less capable of maintaning significant atmospheres.

\begin{figure}
	\centering   
	\includegraphics[width=25pc]{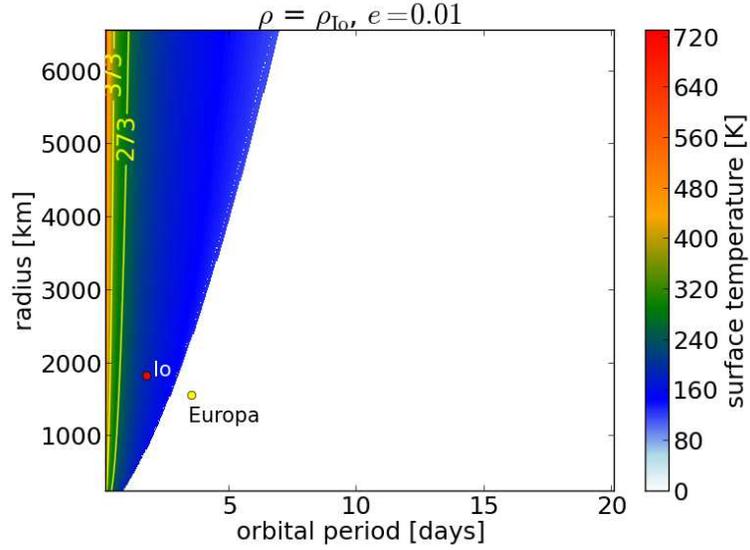}     
	\caption{\label{e001}Tidal heat induced surface temperature for moons with similar density to Io at 0.01 orbital eccentricity.}
\end{figure}

The dependency on the eccentricity can be seen by comparing Fig. \ref{e01} and \ref{e001}. In the case of the latter figure the moon's eccentricity is 0.01. For most of the {\it orbital period--radius} pairs there is no solution (white area). Due to this drastic difference, Europa analogues get out of equilibrium for smaller eccentricities, and the TTZ becomes narrower and shifts to shorter orbital periods. Note, that radiogenic heat is not considered in the model, which could push back the moon into equilibrium state, and would result in higher surface temperature.

\begin{figure}
\centering $
\begin{array}{cc}
\includegraphics[width=19pc]{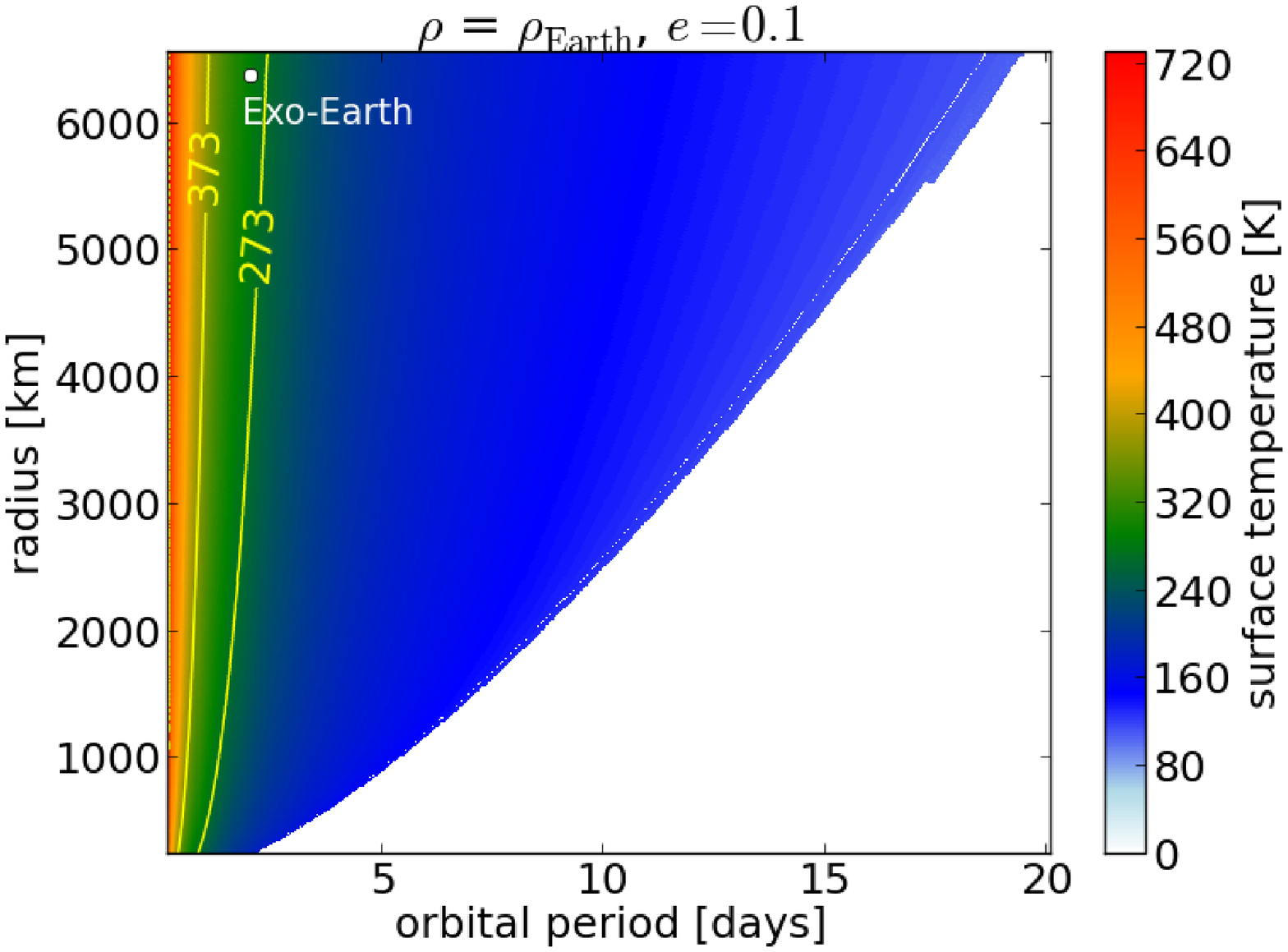} & \includegraphics[width=19pc]{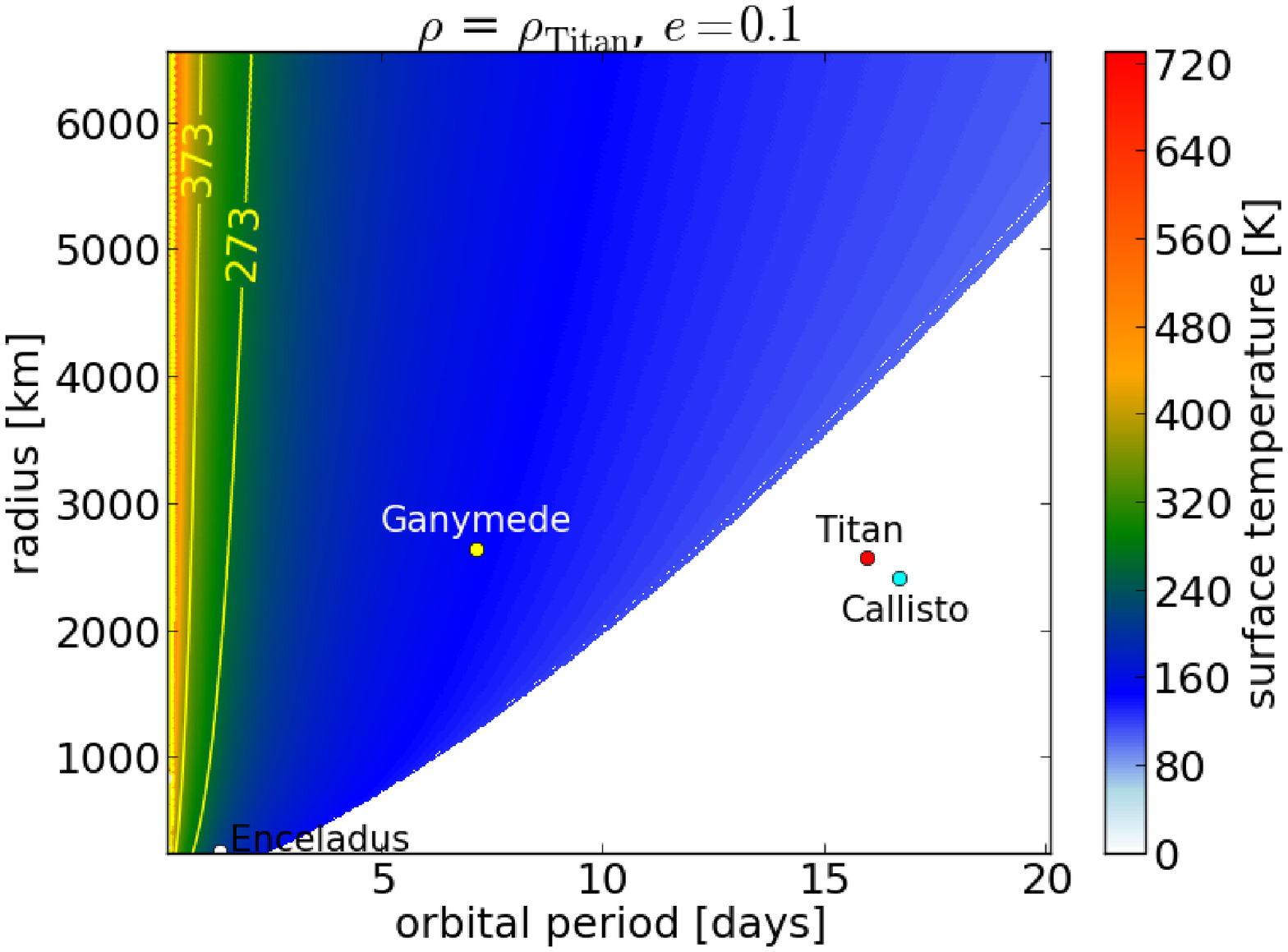} 
\end{array}$
\caption{\label{EarthTitan}Tidal heat induced surface temperature for moons with similar density to the Earth and Titan.}
\end{figure}

Similar calculations were made for the density of the Earth and Titan (left and right panel of Fig. \ref{EarthTitan}, respectively). The densities do not have high influence on the tidally induced surface temperature, however, the TTZ slightly shifts to lower orbital parameters for higher densities. (The density of Earth, Io and Titan are $5515 \, \mathrm{kg/m^3}$, $3528 \, \mathrm{kg/m^3}$ and $1880 \, \mathrm{kg/m^3}$, respectively.)

In the left panel of Fig. \ref{EarthTitan} an example Earth-like moon is plotted inside the TTZ. This hypothetic body has the same mean surface temperature (288~K), radius (6370~km) and density as the Earth, hence its orbital period is 2.06~days. In the right panel a few Solar System satellites are plotted that have similar densities to that of Titan.

The stellar flux for moons with ambient temperatures of $\sim$100~K (which is similar to the case of the Galilean and Saturnian moons in the Solar System) is about one percent of the tidal flux in the TTZ. For this reason, stellar insolation may be safely ignored if the planet-moon system orbits the star at a far distance, or if they are free-floating. For moons system in which the stellar irradiation alone is sufficient to heat the surface to levels of order the melting temperature or higher, the models presented here are would need to be replaced by more complex hybrid ones to take into account both sources of heat and their very different spatial distributions on and within the moon.

\section{Comparison to the fixed $Q$ model}

\subsection{Method}

It is clear from the results, that the viscoelastic model does not give solution in case of small tidal forces. In other words, the amount of heat that is produced by tidal interactions is insufficient to induce convective movements inside the body, and for this reason there is no equilibrium between them. In contrast, the fixed $Q$ model provides solution in these cases, as well. However, the viscoelastic model describes the tidal heating of the body more realistically than the fixed $Q$ model, due to the temperature dependency of the $Q$ and $\mu$ parameters. How are the results of the two models related to each other?

For comparing the results of the two kinds of models, we use the expression of Eq.\,(7) from \citet{peters13} for the fixed $Q$ calculation:

\begin{equation}
   \label{T}
       T_\mathrm{surf} = \left( \left( \frac {392 \pi^5 G^5} {9747 \sigma^2} \right)^{1/2} \left( \frac {R_\mathrm{m}^5 \rho^{9/2}} {\mu Q} \right) \left( \frac {e^2} {\beta^{15/2}} \right) \right)^{1/4} \, ,
\end{equation}

\noindent where $T_\mathrm{surf}$ is the surface temperature of the moon induced by tidal heating, $G$ is the gravitational constant, $\sigma$ is the Stefan-Boltzmann constant, $R_\mathrm{m}$ is the radius, $\rho$ is the densitiy, $\mu$ is the elastic rigidity, and $Q$ is the dissipation function of the moon, $e$ is the eccentricity of the moon's orbit, and $\beta$ is expressed with the semi-major axis ($a$) and the mass of the planet ($M_\mathrm{p}$):
\begin{equation}
   \label{a}
       a = \beta a_R = \beta \left( \frac {3 M_\mathrm{p}} {2 \pi \rho} \right)^{1/3} \, ,
\end{equation}

\noindent where $a_R$ is the Roche radius of the host planet. These equations can be used for calculating the surface temperature of the moon heated solely by tidal forces.

The viscoelastic model is described in details in Sections 2.2 and 2.3. The satellite's mean motion can be expressed from $\beta$ by

\begin{equation}
   \label{n}
       n = \sqrt {\frac {2 \pi G} {3} \frac {\rho} {\beta^3} } \, ,
\end{equation}

\noindent which makes the comparison of the two models easier.

\subsection{Surface temperature}

\begin{figure}
	\centering   
	\includegraphics[width=25pc]{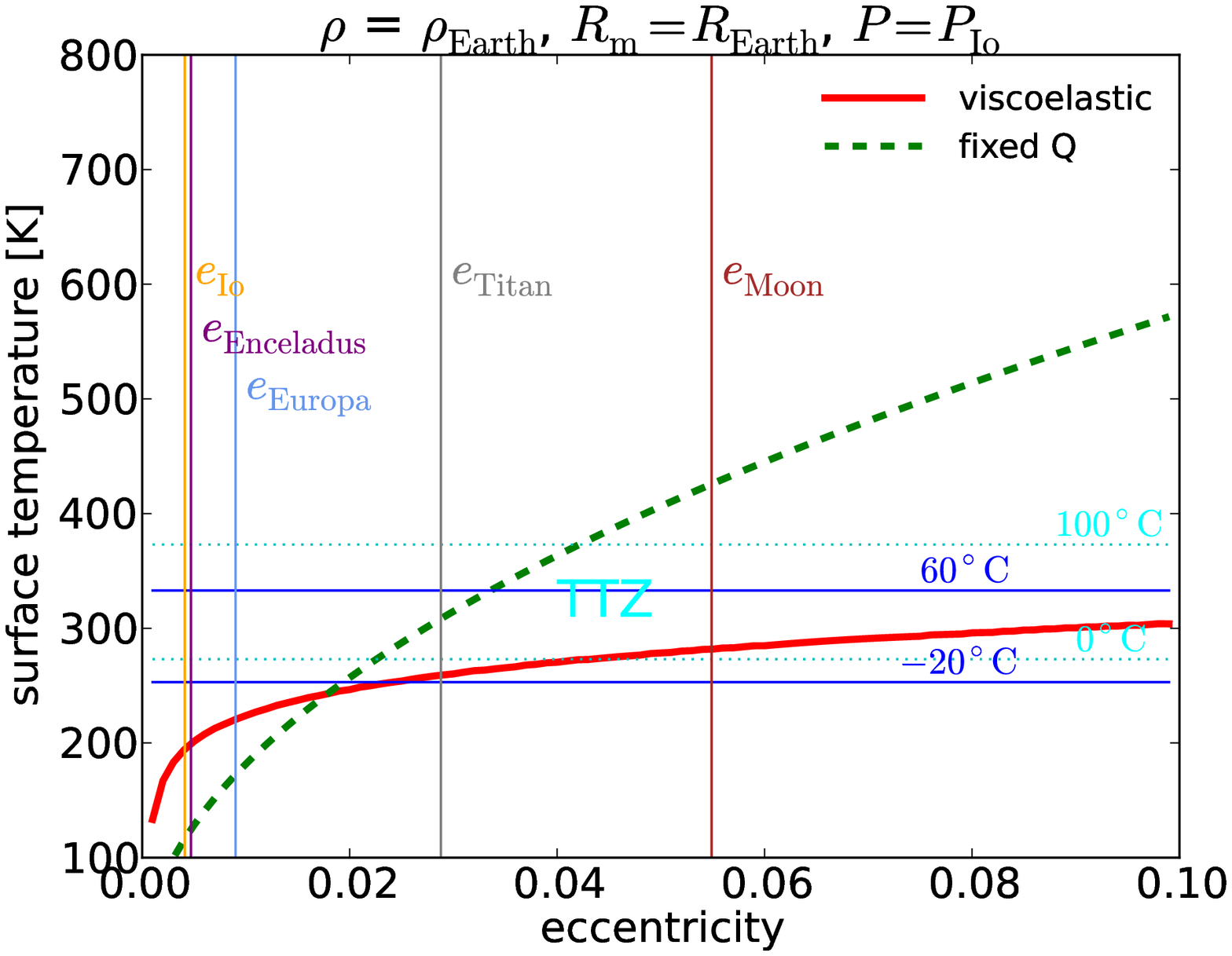}     
	\caption{\label{T-e}Tidally induced surface temperature of the satellite as fuction of its eccentricity.}
\end{figure}

\begin{figure}
\centering   
\includegraphics[width=25pc]{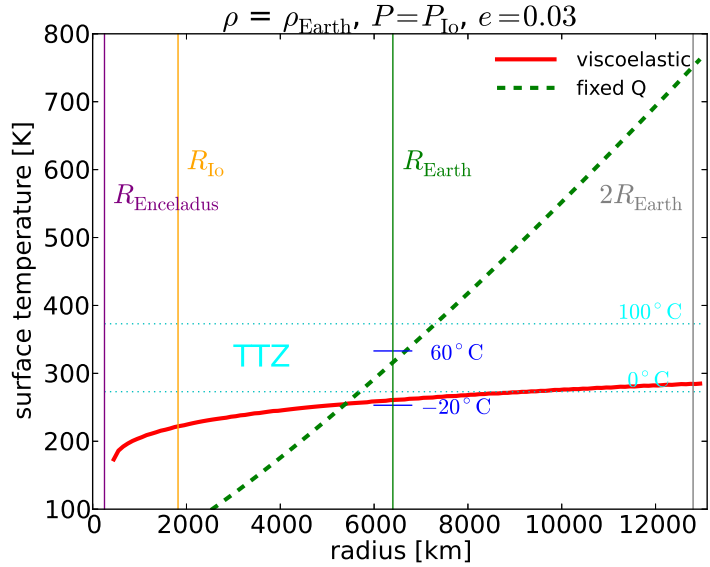}     
\caption{\label{T-R}Tidally induced surface temperature of the satellite as fuction of its radius.}
\end{figure}

\begin{figure}
\centering   
\includegraphics[width=25pc]{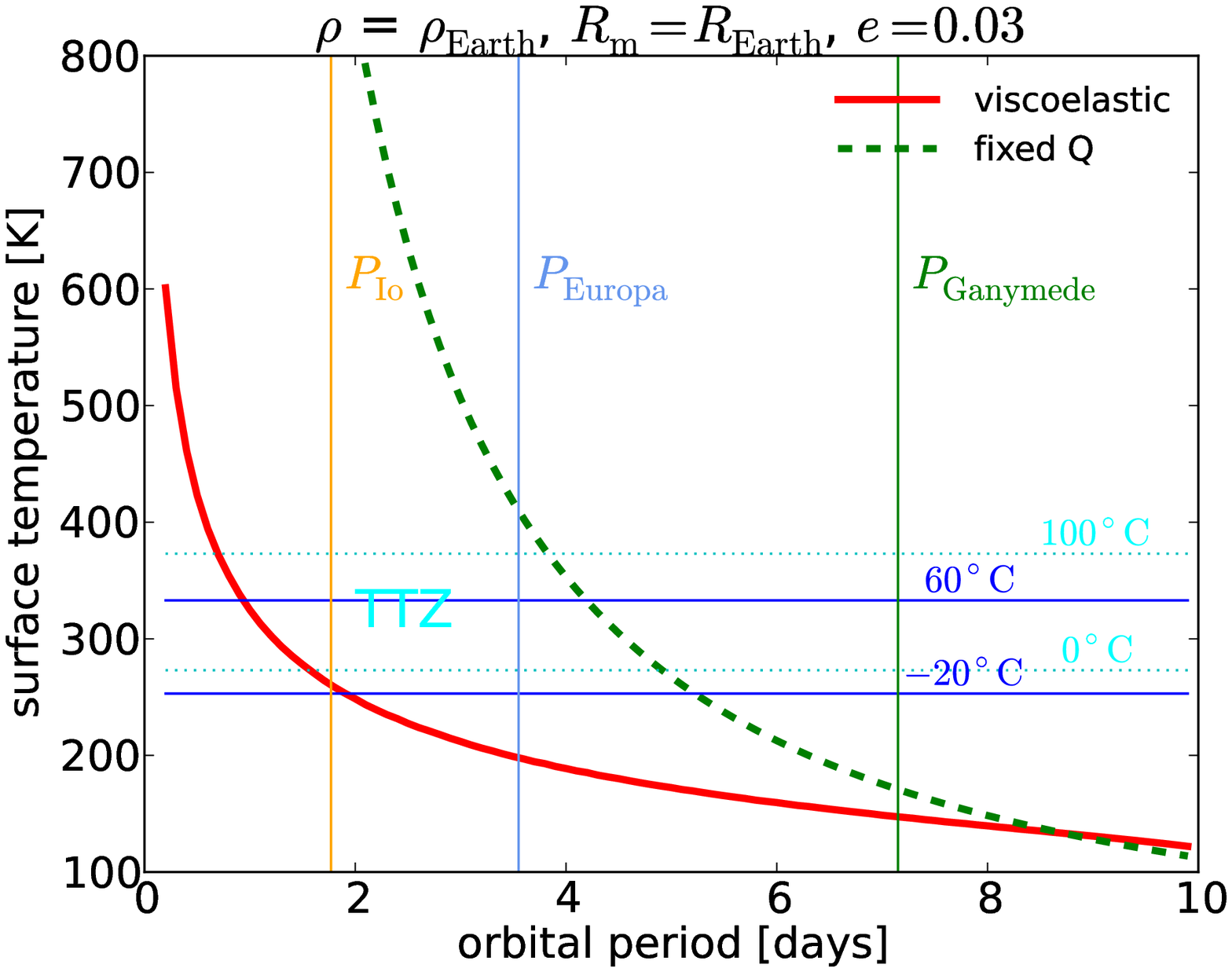}     
\caption{\label{T-P}Tidally induced surface temperature of the satellite as fuction of its orbital period.}
\end{figure}

For comparison of the fixed $Q$ and the viscoelastic model, see Figs. \ref{T-e}, \ref{T-R} and \ref{T-P}, which show the surface temperature of a moon, calculated with both the viscoelastic (red solid curve) and the fixed $Q$ model (green dashed curve) as functions of the eccentricity, radius and orbital period of the satellite. For the density of the moon we used 5515~$\mathrm{kg/m^3}$, which is the density of the Earth, and for the fixed $Q$ model we used $Q=280$ and $\mu = 12 \cdot 10^{10} \, \mathrm{kg / ( m \, s^2 ) }$ in each case \citep[Table 1]{peters13}. The radius, orbital period and eccentricity of the satellite are set to that of the Earth, Io and 0.03, respectively, except that one of these parameters is varied in each figures (horizontal axes). The horizontal light blue, dashed lines indicate 0 and 100~$^\circ$C (making the boundaries of the TTZ), and the solid blue lines denote the minimum and maximum temperatures ($-20$ and $60\,^\circ$C) that are probable limits of habitability on an Earth-like body \citep[][Chapter 4]{sullivan07}. In salty solutions the lower limit for microbial activity is around $-20\,^\circ$C, and the upper limit for complex eukaryotic life is $60\,^\circ$C. The latter temperature is also about the runaway greenhouse limit for Earth. These limits are only used for Earth-like bodies ($\rho = \rho_\mathrm{Earth}$ and $R_\mathrm{m} \approx R_\mathrm{Earth}$). Vertical lines show a few examples from the Solar System for different eccentricities, radii and orbital periods.

It is noticeable that the red curve is less steep than the green curve, and larger portion of it is located inside the TTZ, especially in Figs.~\ref{T-e} and \ref{T-R}. It shows that the viscoelastic model stabilizes the surface temperature comparing to the fixed $Q$ model. These are just a few examples indicating that the viscoelastic model is less sensitive to these parameters, and that there are huge differences in the results of the models. In the next section the volume of the TTZ is investigated more thoroughly.

\subsection{Occurrence rate of 'habitable' moons}

\begin{figure}
	\centering   
	\includegraphics[width=25pc]{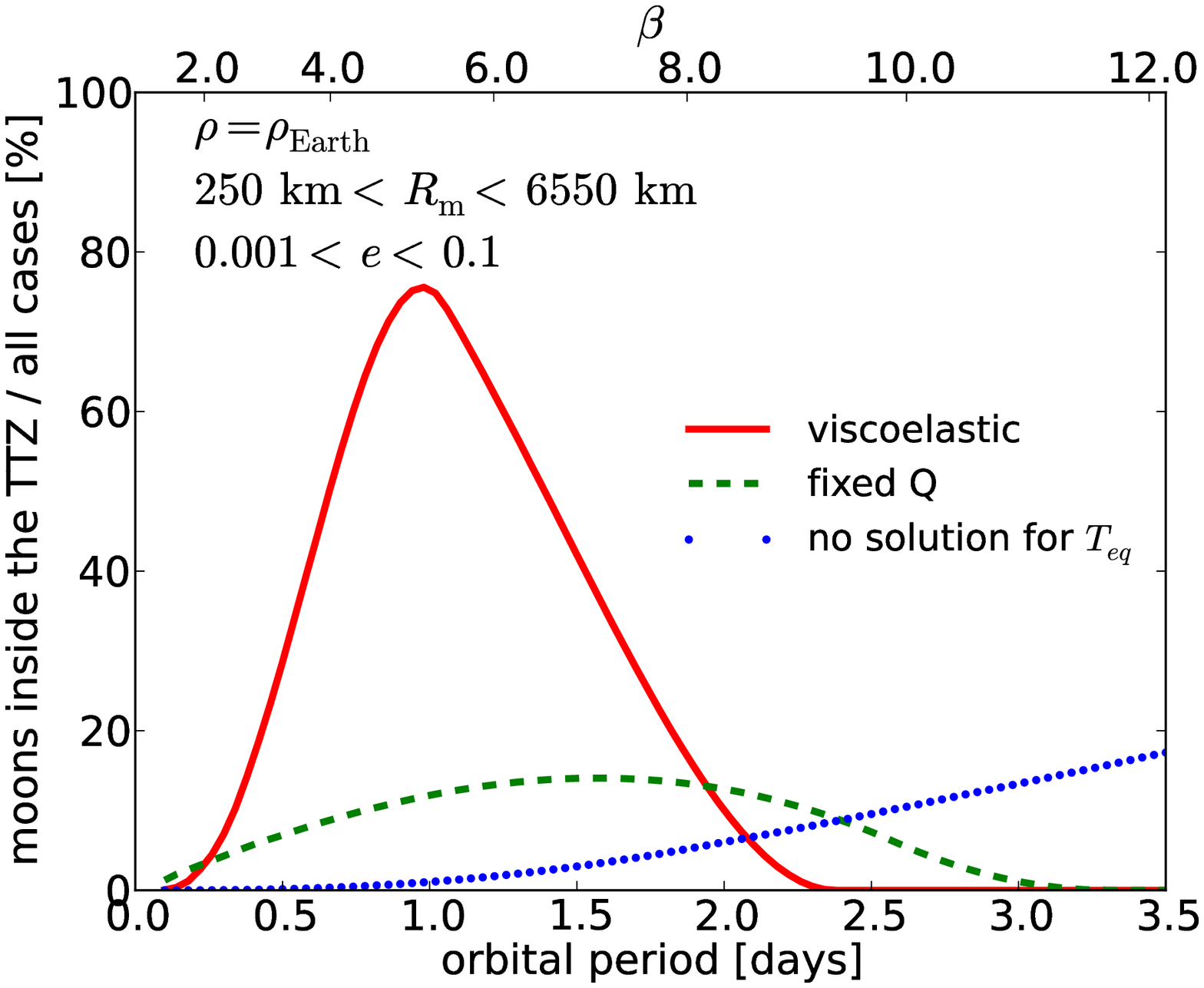}     
	\caption{\label{beta}Percentage of cases that give surface temperatures between 0 and 100~$^\circ$C. Solid red and dashed green curves represent the results of the viscoelastic and the fixed $Q$ model, respectively. The density of the moon is that of the Earth in each case. The dotted blue curve shows the ratio in percentage of those cases where the viscoelastic model did not have solution for the equilibrium temperature.}
\end{figure}

Habitability on extraterrestrial bodies is an exciting, but complex question. Here we consider solely the tidally induced surface temperature of a hypothetic moon. We were curious about the occurrence rate of moons with suitable surface temperature for life. For this reason we mapped the phase space evenly with hypothetic moons that have different radii (between 250 and 6550~km) and eccentricities (between 0.001 and 0.1), and their densities are that of the Earth. We used both the viscoelastic and the fixed $Q$ model for calculating the surface temperature of these bodies, and then calculated the percentage of those that have suitable surface temperature, i.e. that are located inside the TTZ ($0 \leq T_\mathrm{surf} \leq 100~^\circ$C). The calculation was made for different orbital periods between 0.1 to 3.5~days, and for each value there were 63100 hypothetical moons distributed in the radius-eccentricity phase space. The result can be seen in Fig. \ref{beta}. Red solid and green dashed curves indicate the percentage of being inside the TTZ for the viscoelastic and for the fixed Q model, respectively. The blue dotted curve shows the percentage of those cases that do not give result for the viscoelastic model. The top axis shows the $\beta$ parameter, which is the ratio of the moon's semi-major axis and the planet's Roche-radius. It can be clearly seen that the red and green curves have a peak, which means that the probability of having suitable surface temperature has a maximum at a certain orbital period. The viscoelastic model predicts a much more efficient heating than the fixed $Q$ model, i.e. a much larger fraction of the hypothetical moons have their surface temperature between 0 and 100~$^\circ$C. The ratio of the integral under the red to that under the green curve is 2.8, meaning that 2.8 times more exomoons are predicted in the TTZ with the viscoelastic model. For the viscoelastic model the maximum percentage appears around 1~day orbital period, and here the probability for the moon of being inside the TTZ is almost 80\%. For higher orbital periods, this probability rapidly falls down, which is in contrast with the fixed $Q$ model. The latter has its peak around 1.5~days, and it has less than 20\% chance for satellites being in the TTZ. Despite of the high probabilities achieved by the viscoelastic model for small orbital periods, the fixed $Q$ model give more promising results for those moons that have their orbital periods at 2~days or more.

\begin{figure}
	\centering   
	\includegraphics[width=25pc]{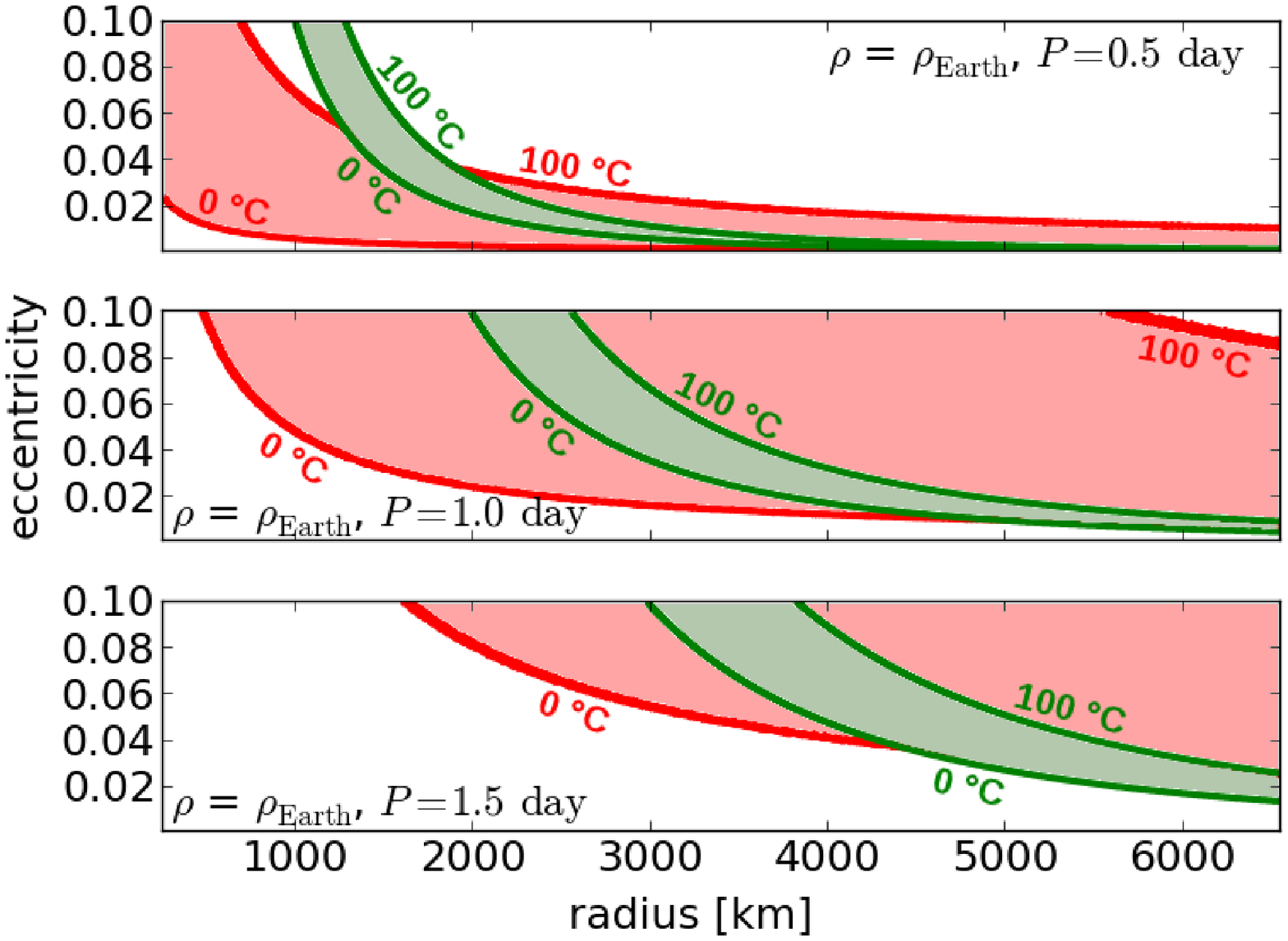}     
	\caption{\label{slice}Temperature contours for the two kinds of models. Red colour: viscoelastic model, green: fixed $Q$ model. Top panel: orbital period $P = 0.5$~day, middle: $P = 1$~day, bottom: $P = 1.5$~days.}
\end{figure}

For detailed study, the 0 and 100~$^\circ$C temperature contours were plotted in the radius-eccentricity plane for a few, specific orbital periods, namely $P = 0.5$~day (top panel), $P = 1$~day (middle panel) and $P = 1.5$~days (bottom panel) (see Fig. \ref{slice}). Again, red and green colours represent the viscoelastic and the fixed $Q$ model, respectively. Between the contour curves the region of the TTZ is filled with light red and light green colours. The result shows that the viscoelastic model mostly favours the small moons, especially at high eccentricities, but also some large moons at small eccentricities over the fixed $Q$ model. This suggests that the viscoelastic model is less sensitive to the parameters of the moon, and holds the temperature more steady than the fixed $Q$ model. This is due to the melting of the inner material of the moon that leads to a less elevated temperature, as discussed by \citet{peters13}. On the other hand, the lower temperature implies that the total irradiated flux of the moon will be also lower, hence making the detection of the moon more difficult.

\begin{figure}
	\centering   
	\includegraphics[width=25pc]{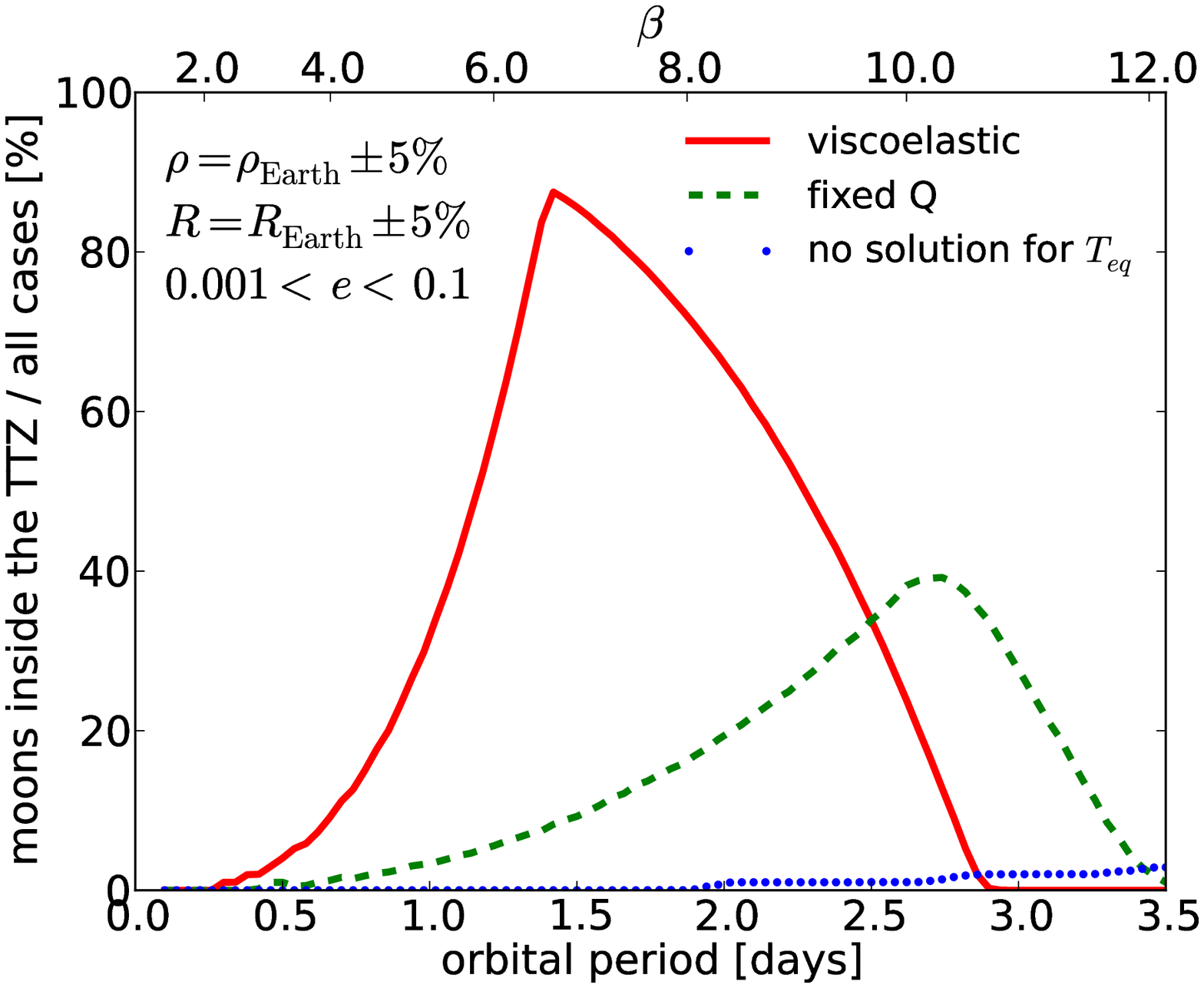}     
	\caption{\label{betaEarth}Percentage of cases that give surface temperatures between $-20$ and 60~$^\circ$C. Solid red and dashed green curves represent the results of the viscoelastic and the fixed Q model, respectively. The density of the moon is that of the Earth in each case. The dotted blue curve shows the ratio in percentage of those cases where the viscoelastic model did not have solution for the equilibrium temperature.}
\end{figure}

We were also interested in 'Earth-twins' as satellites and in the probability of their 'habitability'. For this reason we made similar calculations, but the radius and the density of the hypothetical moons were set to be close to that of the Earth: $R_\mathrm{m} = 6378\, \mathrm{km} (\pm 5\%)$ and  $\rho = 5515\, \mathrm{kg/m^3} (\pm 5\%)$. The radius and density values were chosen randomly from these intervals. The eccentricity was altered similarly as in the previous case (uniformly between 0.001 and 0.1). Altogether 200000 cases were considered for each orbital period. The temperature limits were set to $-20$ and 60~$^\circ$C, which are the probable limits for life on Earth. Fig. \ref{betaEarth} shows the results of this calculation. Note, that the peaks of the red solid (viscoelastic model) and the green dashed (fixed $Q$ model) curves are shifted to higher orbital periods, comparing to Fig. \ref{beta}. This is in part caused by the changed temperature limits, and in part by the much shorter radius range. The maximum probabilities are also higher, that is especially visible in the case of the fixed $Q$ model, which reaches more than 40~\% at the curve's peak (in the previous case it was less that 20~\%). As one would expect, it suggests that larger moons are more probable of maintaining warm surfaces. The ratio of the areas under the red and the green curves is 2.3.

In general, it can be concluded that the viscoelastic model is not just more realistic than the fixed $Q$ model, but also gives more promising results for exomoons, since much larger fraction of the hypothetical satellites have been found in the TTZ. In those cases when the viscoelastic model does not give solution for the equilibrium temperature, one can use the fixed $Q$ model instead, however, the values of $Q$ and $\mu$ are highly uncertain.

\subsection{The value of $Q \mu$}

With the product of $Q$ and $\mu$, one can easily calculate the tidally induced surface temperature of a moon without using a complex viscoelastic model. Using Eq.\,\ref{T} is a fast way to obtain $T_\mathrm{surf}$, but a good approximation is needed for the $Q \mu$ value. For such calculations, in the following, we give the $Q \mu$ values for hypothetical moons. Because of the large number of possible variations in the physical and orbital parameters of the moons, only a few, Solar System-like bodies were considered. Since the $Q \mu$ varies several orders of magnitudes for different rocky bodies, a good estimation can serve almost as well as the exact value. The following examples can be used as a guideline for making such estimations. Note, that the used model can be applied to rocky bodies, but for icy satellites, such as Enceladus or Europa, the results may be misleading, because of the more complex structure and different behaviour of icy material.

\begin{table}
\begin{tabular}{l|c|c|c|c|c|c}
label & $R_\mathrm{m} \, \mathrm{[km]}$ & $\rho \, \mathrm{[kg/m^3]}$ & $e$ & $P \, \mathrm{[days]}$ & $T_\mathrm{surf} \, \mathrm{[K]}$ & $\mathrm{log_{10}}(Q \mu)$ [Pa] \\
\hline
\multirow{6}{*}{Earth-like} & \multirow{6}{*}{6378} & \multirow{6}{*}{5515} & \multirow{3}{*}{0.01} & 1 & 281 & 14.0 \\
 & & & & 2 & 213 & 13.0 \\
 & & & & 3 & 180 & 12.4 \\
\cline{4-7}
 & & & \multirow{3}{*}{0.1} & 1 & 378 & 15.5 \\
 & & & & 2 & 291 & 14.4 \\
 & & & & 3 & 249 & 13.8 \\
\hline
\multirow{6}{*}{Mars-like} & \multirow{6}{*}{3394} & \multirow{6}{*}{3933} & \multirow{3}{*}{0.01} & 1 & 256 & 12.5 \\
 & & & & 2 & 194 & 11.5 \\
 & & & & 3 & 163 & 10.9 \\
\cline{4-7}
 & & & \multirow{3}{*}{0.1} & 1 & 342 & 14.0 \\
 & & & & 2 & 263 & 13.0 \\
 & & & & 3 & 225 & 12.4 \\
\hline
\multirow{6}{*}{Moon-like} & \multirow{6}{*}{1738} & \multirow{6}{*}{3342} & \multirow{3}{*}{0.01} & 1 & 232 & 11.1 \\
 & & & & 2 & 176 & 10.1 \\
 & & & & 3 & n.a. & n.a. \\
\cline{4-7}
 & & & \multirow{3}{*}{0.1} & 1 & 313 & 12.6 \\
 & & & & 2 & 240 & 11.5 \\
 & & & & 3 & 205 & 10.9 \\
\hline
\multirow{6}{*}{Io-like} & \multirow{6}{*}{1821} & \multirow{6}{*}{3532} & \multirow{3}{*}{0.01} & 1 & 235 & 11.2 \\
 & & & & 2 & 177 & 10.2 \\
 & & & & 3 & n.a. & n.a. \\
\cline{4-7}
 & & & \multirow{3}{*}{0.1} & 1 & 315 & 12.7 \\
 & & & & 2 & 242 & 11.7 \\
 & & & & 3 & 207 & 11.1 \\
\end{tabular}
\caption{\label{Qmu}$Q \mu$ values for Solar System-like, rocky moons. The radii and densities are those of the corresponding Solar System bodies. The reference for these values is \citet[Appendix A]{murray99}.}
\end{table}

From the surface temperature of the moon, that was calculated from the viscoelastic model, the $Q \mu$ value was determined using Eq.\,\ref{T} for six orbital period--eccentricity pairs. In Table \ref{Qmu} the tidally induced surface temperatures and the logarithm of the $Q \mu$ values can be seen. The radius and density of the satellites are those of the corresponding Solar System bodies (see the first column), and the values are from \citet{murray99}. The eccentricities are set to 0.01 and 0.1, and the orbital periods to 1, 2 and 3 days. 'N.a.' indicates that there was no solution (weak tidal forces).

\subsection{Scaling the Galilean satellite system}

Since no satellite has been discovered so far outside the Solar System, we used the Galilean system as a prototype for realistic calculations. Io, Europa and Ganymede are orbiting in a 1:2:4 mean motion resonance, that maintains their eccentricities, which play an essential role in forcing continuously their tidal heating. \citet{ogihara12} investigated satellite formation in the circumplanetary disc of giant planets using $N$-body simulations including gravitational interactions with the circumplanetary gas disc. They have found that 2:1 mean motion resonances are almost inevitable in Galilean-like satellite systems, and based on their results they predict that mean motion resonances may be common in exoplanetary systems. For these reasons the Galilean satellite system seems to be a representative example for realistic calculations, since the moons are in resonance, and their scaled-up versions will probably stay in resonance, too.

The test systems consist of a planet (Jupiter) and the four Galilean moons. 91 cases are considered, one is the real Galilean system, and the others are the scaled-up versions: the masses of the planet and the moons were multiplied by the $scale$ factor ($scale = 1.0, 1.1, 1.2, ... \, 10.0$), and the semimajor axes of the moons were altered with constant orbital periods for each $scale$ value.

\begin{equation}
   \label{P}
       P = 2 \pi \sqrt { \frac {a^3} {scale \cdot G (M_\mathrm{p} + M_\mathrm{s})} } \, ,
\end{equation}

\noindent where $a$ is the semimajor axis of the moon, $M_\mathrm{p}$ and $M_\mathrm{s}$ are the masses of the planet and the moon, respectively. The fixed orbital periods guarantee that the satellites approximately stay in resonances. This calculation resulted in constant $\beta$ values for all $scale$ parameters.

\begin{figure}
\centering   
\includegraphics[width=25pc]{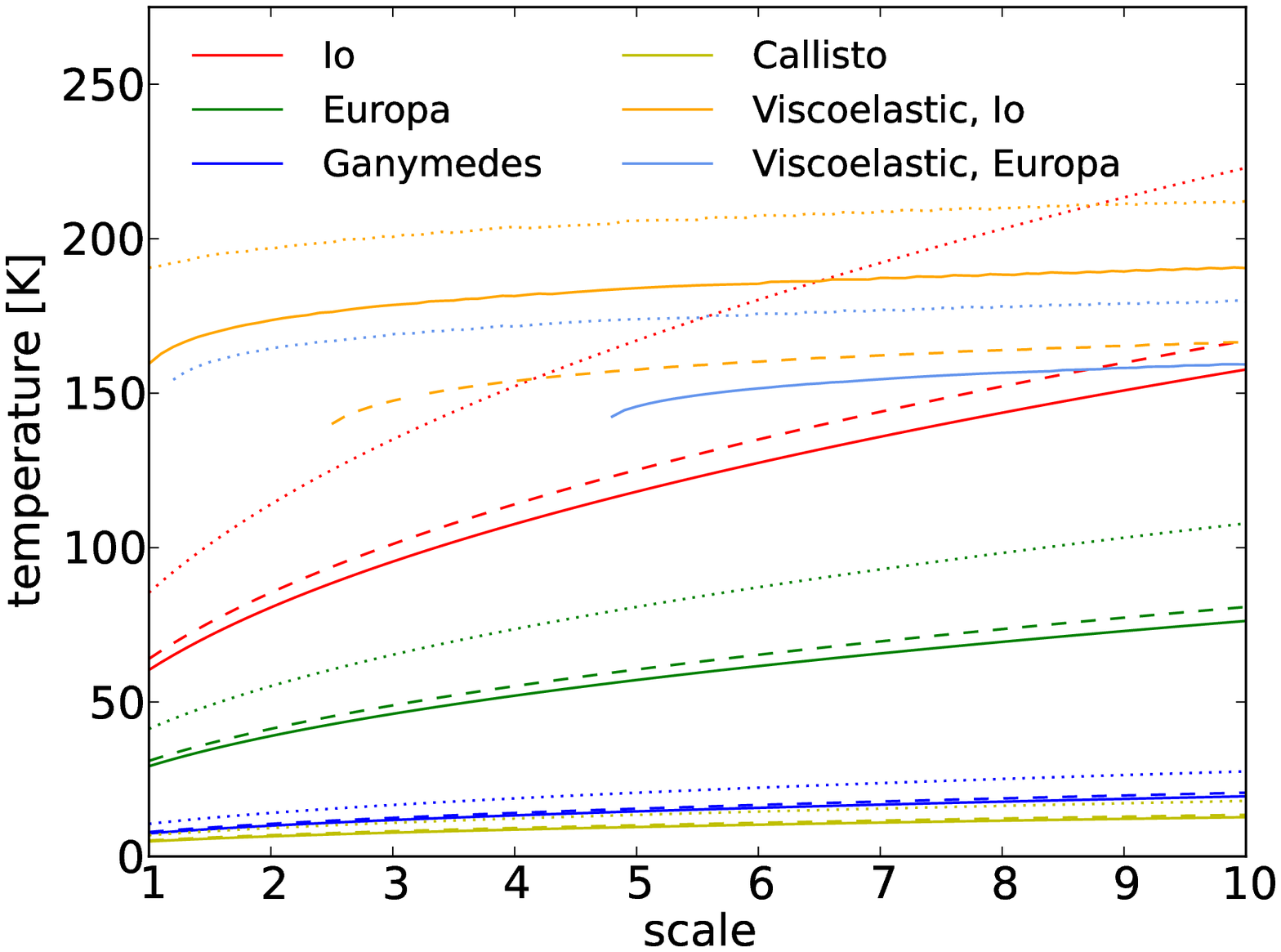}
\caption{\label{scaleJup}Surface temperature of the Galilean moons as function of the $scale$ parameter, calculated by both the fixed $Q$ and the viscoelastic model. Dashed and dotted curves indicate that the density or the eccentricity of the moon is doubled compared to the Solar System case, respectively.}
\end{figure}

Using both the fixed $Q$ and the viscoelastic models, the warmth of tidal heat was investigated in each case. The tidal heat induced surface temperature can be seen in Fig. \ref{scaleJup}, where the 91 cases are connected with solid curves for each satellite. 182 other cases were calculated, too, which are shown with dashed and dotted curves in the figure. These curves indicate that the densities (dashed curve) and the eccentricities (dotted curves) of the satellites are doubled compared to their original values in the Solar System. In the calculations $\mu$ and $Q$ were set for all satellites to that of Io, namely $10^{10} \mathrm{\,kg / (m \, s^2} )$ and 36, respectively, except for Europa, which has the following parameters: $\mu = 4 \cdot 10^{9} \mathrm{\,kg / (m \, s^2} )$ and $Q = 100$ \citep{peters13}. Densities of the moons are from \citet{lodders98}, and the reference for the semi-major axis, eccentricity and mass values is \citet[Appendix A]{murray99}.

For Io, in the $scale = 1$ (Solar System) case the fixed $Q$ and the viscoelastic models give 60~K and 160~K, respectively. The observed surface heat flux induced by tidal heat on Io is around $2 \mathrm{\,W / m^2}$, which is a lower limit \citep{spencer00}. In other words, tidal forces produce at least 77~K heat on the surface of Io. The fixed $Q$ model resulted in a lower value than this limit, but note that $Q$ and $\mu$ are very difficult to estimate. The viscoelastic model gave much higher temperature than the observation, but keep in mind, that the heat is concentrated in hotspots, and is not evenly distributed on the surface of Io. The temperature of the most warm volcano, Loki is higher than 300~K \citep{spencer00}.

The viscoelastic model gives solution only for Io (orange curves) and Europa (light blue curves), but not for all $scale$ values, as shown in Fig.~\ref{scaleJup}. In those cases when the densities are the twice than those in the Solar System (dashed curves), the surface temperatures are just slightly higher. In fact, in the viscoelastic model, higher densities result in less tidal heat because of the imaginary part of the second order Love number. Doubling the eccentricity instead of the density (dotted curves) makes the surface temperature higher in each case.

\section{Conclusions}

We have used, for the first time, a viscoelastic model for calculating the surface temperature of tidally heated exomoons. The viscoelastic model gives more reliable results than the widely used fixed $Q$ model, because it takes into account that the tidal dissipation factor ($Q$) and rigidity ($\mu$) strongly depend on the temperature. Besides, these values are poorly known even for planets and satellites in the Solar System. Using the viscoelastic model for exomoons helps to get a more realistic estimation of their surface temperature, and to determine a circumplanetary region, where liquid water may exist on them. It may help future missions in selecting targets for exomoon detections.

We have defined the Tidal Temperate Zone, which is the region around a planet where the surface temperature of the satellite is between $0~^\circ$C and $100~^\circ$C. No sources of heat were considered other than tidal forces. Assuming, that the planet-moon system orbits the star at a far distance, or the stellar radiation is low due to the spectral type, tidal heat can be the dominant heat source affecting the satellite. We have investigated such systems, and found that the TTZ strongly depends on the orbital period, and less on the radius of the moon. For higher densities or eccentricities of the moon, the location of the TTZ is slightly closer to the planet.

Comparing this model to the traditionally used fixed $Q$ model revealed that there are huge differences in the results. Generally, the viscoelastic model is less sensitive to moon radius than the uniform $Q$ model, keeping the surface temperature of the body more steady. The reason is that higher tidal forces induce higher melt fraction which results in a lower temperature than the fixed $Q$ model. The viscoelastic model demonstrates the way in which partially melting of a moon can act as a thermostat and tend to fix its temperature somewhere near its melting point over a wide range of physical and orbital parameters. As a consequence, the statistic volume of the TTZ is much larger in the viscoelastic case, which is favourable for life. But this lower temperature also means that the detectability of such moons is lower in the infrared. In addition, for low tidal forces there is no equilibrium with the convective cooling; hence, only the fixed $Q$ model provides solution. In these cases the challenge is to determine the values of $Q$ and $\mu$.

For a few characteristic cases the product of the tidal dissipation factor and rigidity was calculated from the viscoelastic model, in order to help in quick estimations of tidally heated exomoon surface temperatures. Since the viscoelastic model is more realistic because of the inner melting and the temperature dependency of the parameters, but the fixed $Q$ model is easier to use, these $Q \mu$ values (along with the surface temperature) are provided in Table \ref{Qmu}. By inserting $Q \mu$ into Eq.\,\ref{T}, one can get the estimation of the tidally induced surface temperature of a moon. Connection between the quality factor ($Q$) and the viscoelastic parameters (viscosity and shear modulus) was given for the Maxwell model, too, by \citet{remus12}.

Earth-like bodies were also investigated as satellites, and in these cases the $-20$ and $60~^\circ$C temperatures were used as limits of habitability. The results are similar, but the volume of this habitable zone is larger than that of the TTZ for wide range of satellite radii. This habitable zone includes atmospheric considerations of the moon, but stellar radiation was neglected in the calculations. In case of significant radiation from other sources, the surface temperature of the moon will be higher. Additional heat sources (such as stellar radiation, radiogenic processes, reflected stellar and emitted thermal radiation from the planet), and the effects of eclipses, or the obliquity of the satellite are thoroughly discussed by \citet{heller13}. 

For simulating realistic systems, the Galilean moons were used as a prototype. Their surface temperature was calculated with both models for different, scaled up masses. The mean motion resonance between the satellites helps to maintain their eccentricity, and consequently to maintain the tidal forces. By raising their masses, the temperatures of Io and Europa elevates less drastically in the viscoelastic model, than in the fixed $Q$ model (see Fig. \ref{scaleJup}). At $scale = 5$ (masses are five times as in the Solar System case) the surface temperature of Europa is $\sim150$~K calculated from the viscoelastic model. Assuming that its density does not change, its radius will be approximately 0.25 Earth radii. In case of an additional 100--120~K heat (e. g. from stellar radiation), the ice would melt, and this super-Europa would become an 'ocean moon', covered entirely by global water ocean. The used viscoelastic model might not be adequate, and can be oversimplified for such bodies that consist of rocky and icy layers, as well. Salty ice mixtures may also modify the results. The applied model ignores the structure, pressure and other effects, and applies melting for the whole body. However, it provides a global picture of the tidally heated moon. Even with a more detailed viscoelastic model, that describes Enceladus as a three layered body (rocky core, ocean and ice shell), \citet{barr08} have found that tidal heat is $\sim10$ times lower than that was observed by the Cassini Composite Infrared Spectrometer. Similarly, \citet{moore03} concluded that observed heat flux on Io is about an order of magnitude higher than that can be explained with a multilayered, viscoelastic model. These results suggest that tidal heat can be much more relevant than what is predicted by models.

\acknowledgments
We thank Amy Barr, Ren\'e Heller, Edwin Kite and Omer Bromberg for useful conversations. We are also grateful to L\'aszl\'o L. Kiss and Darren Williams for useful comments that greatly improved the manuscript. This research has been supported in part by the World Premier International Research Center Initiative, MEXT, Japan. VD has been supported by the Hungarian OTKA Grants K83790, K104607 and the Lend\"ulet-2009 Young Researchers Program of the Hungarian Academy of Sciences, and the ESA PECS Contract No. 4000110889/14/NL/NDe.

\bibliographystyle{apj}
\bibliography{ref}

\begin{thebibliography}{}
\expandafter\ifx\csname natexlab\endcsname\relax\def\natexlab#1{#1}\fi

\bibitem[{{Barr}(2008)}]{barr08}
{Barr}, A.~C. 2008, Journal of Geophysical Research (Planets), 113, 7009

\bibitem[{{Bennett} {et~al.}(2014){Bennett}, {Batista}, {Bond}, {Bennett},
  {Suzuki}, {Beaulieu}, {Udalski}, {Donatowicz}, {Bozza}, {Abe}, {Botzler},
  {Freeman}, {Fukunaga}, {Fukui}, {Itow}, {Koshimoto}, {Ling}, {Masuda},
  {Matsubara}, {Muraki}, {Namba}, {Ohnishi}, {Rattenbury}, {Saito}, {Sullivan},
  {Sumi}, {Sweatman}, {Tristram}, {Tsurumi}, {Wada}, {Yock}, {The MOA
  Collaboration}, {Albrow}, {Bachelet}, {Brillant}, {Caldwell}, {Cassan},
  {Cole}, {Corrales}, {Coutures}, {Dieters}, {Dominis Prester}, {Fouqu{\'e}},
  {Greenhill}, {Horne}, {Koo}, {Kubas}, {Marquette}, {Martin}, {Menzies},
  {Sahu}, {Wambsganss}, {Williams}, {Zub}, {The PLANET Collaboration}, {Choi},
  {DePoy}, {Dong}, {Gaudi}, {Gould}, {Han}, {Henderson}, {McGregor}, {Lee},
  {Pogge}, {Shin}, {Yee}, {The {$\mu$}FUN Collaboration}, {Szyma{\'n}ski},
  {Skowron}, {Poleski}, {Koz{\l}owski}, {Wyrzykowski}, {Kubiak},
  {Pietrukowicz}, {Pietrzy{\'n}ski}, {Soszy{\'n}ski}, {Ulaczyk}, {The OGLE
  Collaboration}, {Tsapras}, {Street}, {Dominik}, {Bramich}, {Browne},
  {Hundertmark}, {Kains}, {Snodgrass}, {Steele}, {The RoboNet Collaboration},
  {Dekany}, {Gonzalez}, {Heyrovsk{\'y}}, {Kandori}, {Kerins}, {Lucas},
  {Minniti}, {Nagayama}, {Rejkuba}, {Robin}, \& {Saito}}]{bennett14}
{Bennett}, D.~P., {Batista}, V., {Bond}, I.~A., {et~al.} 2014, \apj, 785, 155

\bibitem[{{Canup} \& {Ward}(2006)}]{canup06}
{Canup}, R.~M., \& {Ward}, W.~R. 2006, \nat, 441, 834

\bibitem[{{Carr} {et~al.}(1998){Carr}, {Belton}, {Chapman}, {Davies},
  {Geissler}, {Greenberg}, {McEwen}, {Tufts}, {Greeley}, {Sullivan}, {Head},
  {Pappalardo}, {Klaasen}, {Johnson}, {Kaufman}, {Senske}, {Moore}, {Neukum},
  {Schubert}, {Burns}, {Thomas}, \& {Veverka}}]{carr98}
{Carr}, M.~H., {Belton}, M.~J.~S., {Chapman}, C.~R., {et~al.} 1998, \nat, 391,
  363

\bibitem[{{Collins} \& {Goodman}(2007)}]{collins07}
{Collins}, G.~C., \& {Goodman}, J.~C. 2007, \icarus, 189, 72

\bibitem[{{Fischer} \& {Spohn}(1990)}]{fischer90}
{Fischer}, H.-J., \& {Spohn}, T. 1990, \icarus, 83, 39

\bibitem[{{Goldreich} \& {Soter}(1966)}]{goldreich66}
{Goldreich}, P., \& {Soter}, S. 1966, \icarus, 5, 375

\bibitem[{{Heller} \& {Barnes}(2013)}]{heller13}
{Heller}, R., \& {Barnes}, R. 2013, Astrobiology, 13, 18

\bibitem[{{Heller} {et~al.}(2014){Heller}, {Williams}, {Kipping}, {Limbach},
  {Turner}, {Greenberg}, {Sasaki}, {Bolmont}, {Grasset}, {Lewis}, {Barnes}, \&
  {Zuluaga}}]{helleretal14}
{Heller}, R., {Williams}, D., {Kipping}, D., {et~al.} 2014, Astrobiology, 14,
  798

\bibitem[{{Henning} {et~al.}(2009){Henning}, {O'Connell}, \&
  {Sasselov}}]{henning09}
{Henning}, W.~G., {O'Connell}, R.~J., \& {Sasselov}, D.~D. 2009, \apj, 707,
  1000

\bibitem[{{Hinkel} \& {Kane}(2013)}]{hinkel13}
{Hinkel}, N.~R., \& {Kane}, S.~R. 2013, \apj, 774, 27

\bibitem[{{Iess} {et~al.}(2014){Iess}, {Stevenson}, {Parisi}, {Hemingway},
  {Jacobson}, {Lunine}, {Nimmo}, {Armstrong}, {Asmar}, {Ducci}, \&
  {Tortora}}]{iess14}
{Iess}, L., {Stevenson}, D.~J., {Parisi}, M., {et~al.} 2014, Science, 344, 78

\bibitem[{{Kaltenegger}(2010)}]{kaltenegger10}
{Kaltenegger}, L. 2010, \apjl, 712, L125

\bibitem[{{Kargel} {et~al.}(2000){Kargel}, {Kaye}, {Head}, {Marion}, {Sassen},
  {Crowley}, {Ballesteros}, {Grant}, \& {Hogenboom}}]{kargel00}
{Kargel}, J.~S., {Kaye}, J.~Z., {Head}, J.~W., {et~al.} 2000, \icarus, 148, 226

\bibitem[{{Kipping}(2009{\natexlab{a}})}]{kipping09a}
{Kipping}, D.~M. 2009{\natexlab{a}}, \mnras, 392, 181

\bibitem[{{Kipping}(2009{\natexlab{b}})}]{kipping09b}
---. 2009{\natexlab{b}}, \mnras, 396, 1797

\bibitem[{{Kipping} {et~al.}(2012){Kipping}, {Bakos}, {Buchhave},
  {Nesvorn{\'y}}, \& {Schmitt}}]{kipping12}
{Kipping}, D.~M., {Bakos}, G.~{\'A}., {Buchhave}, L., {Nesvorn{\'y}}, D., \&
  {Schmitt}, A. 2012, \apj, 750, 115

\bibitem[{{Lodders} \& {Fegley}(1998)}]{lodders98}
{Lodders}, K., \& {Fegley}, B. 1998, {The planetary scientist's companion /
  Katharina Lodders, Bruce Fegley.} (Oxford University Press)

\bibitem[{{Meyer} \& {Wisdom}(2007)}]{meyer07}
{Meyer}, J., \& {Wisdom}, J. 2007, \icarus, 188, 535

\bibitem[{{Moore}(2003)}]{moore03}
{Moore}, W.~B. 2003, Journal of Geophysical Research (Planets), 108, 5096

\bibitem[{{Murray} \& {Dermott}(1999)}]{murray99}
{Murray}, C.~D., \& {Dermott}, S.~F. 1999, {Solar system dynamics} (Cambridge
  University Press)

\bibitem[{{Ogihara} \& {Ida}(2012)}]{ogihara12}
{Ogihara}, M., \& {Ida}, S. 2012, \apj, 753, 60

\bibitem[{{Peters} \& {Turner}(2013)}]{peters13}
{Peters}, M.~A., \& {Turner}, E.~L. 2013, \apj, 769, 98

\bibitem[{{Porco} {et~al.}(2006){Porco}, {Helfenstein}, {Thomas}, {Ingersoll},
  {Wisdom}, {West}, {Neukum}, {Denk}, {Wagner}, {Roatsch}, {Kieffer}, {Turtle},
  {McEwen}, {Johnson}, {Rathbun}, {Veverka}, {Wilson}, {Perry}, {Spitale},
  {Brahic}, {Burns}, {Del Genio}, {Dones}, {Murray}, \& {Squyres}}]{porco06}
{Porco}, C.~C., {Helfenstein}, P., {Thomas}, P.~C., {et~al.} 2006, Science,
  311, 1393

\bibitem[{{Remus} {et~al.}(2012){Remus}, {Mathis}, {Zahn}, \&
  {Lainey}}]{remus12}
{Remus}, F., {Mathis}, S., {Zahn}, J.-P., \& {Lainey}, V. 2012, \aap, 541, A165

\bibitem[{{Reynolds} {et~al.}(1987){Reynolds}, {McKay}, \&
  {Kasting}}]{reynolds87}
{Reynolds}, R.~T., {McKay}, C.~P., \& {Kasting}, J.~F. 1987, Advances in Space
  Research, 7, 125

\bibitem[{{Ross} \& {Schubert}(1988)}]{ross88}
{Ross}, M.~N., \& {Schubert}, G. 1988, Icarus, 78, 90

\bibitem[{{Roth} {et~al.}(2014){Roth}, {Saur}, {Retherford}, {Strobel},
  {Feldman}, {McGrath}, \& {Nimmo}}]{roth14}
{Roth}, L., {Saur}, J., {Retherford}, K.~D., {et~al.} 2014, Science, 343, 171

\bibitem[{{Scharf}(2006)}]{scharf06}
{Scharf}, C.~A. 2006, \apj, 648, 1196

\bibitem[{{Segatz} {et~al.}(1988){Segatz}, {Spohn}, {Ross}, \&
  {Schubert}}]{segatz88}
{Segatz}, M., {Spohn}, T., {Ross}, M.~N., \& {Schubert}, G. 1988, \icarus, 75,
  187

\bibitem[{{Shoji} \& {Kurita}(2014)}]{shoji14}
{Shoji}, D., \& {Kurita}, K. 2014, \apj, 789, 3

\bibitem[{{Simon} {et~al.}(2007){Simon}, {Szatm{\'a}ry}, \&
  {Szab{\'o}}}]{simon07}
{Simon}, A., {Szatm{\'a}ry}, K., \& {Szab{\'o}}, G.~M. 2007, \aap, 470, 727

\bibitem[{{Simon} {et~al.}(2012){Simon}, {Szab{\'o}}, {Kiss}, \&
  {Szatm{\'a}ry}}]{simon12}
{Simon}, A.~E., {Szab{\'o}}, G.~M., {Kiss}, L.~L., \& {Szatm{\'a}ry}, K. 2012,
  \mnras, 419, 164

\bibitem[{{Simon} {et~al.}(2010){Simon}, {Szab{\'o}}, {Szatm{\'a}ry}, \&
  {Kiss}}]{simon10}
{Simon}, A.~E., {Szab{\'o}}, G.~M., {Szatm{\'a}ry}, K., \& {Kiss}, L.~L. 2010,
  \mnras, 406, 2038

\bibitem[{{Spencer} {et~al.}(2000){Spencer}, {Rathbun}, {Travis}, {Tamppari},
  {Barnard}, {Martin}, \& {McEwen}}]{spencer00}
{Spencer}, J.~R., {Rathbun}, J.~A., {Travis}, L.~D., {et~al.} 2000, Science,
  288, 1198

\bibitem[{{Sullivan} \& {Baross}(2007)}]{sullivan07}
{Sullivan}, III, W.~T., \& {Baross}, J. 2007, {Planets and Life} (Cambridge
  University Press)

\bibitem[{{Williams}(2013)}]{williams13}
{Williams}, D.~M. 2013, Astrobiology, 13, 315

\end{thebibliography}
\end{document}